\documentclass[10pt,twocolumn,british,english]{article}
\usepackage[utf8]{luainputenc}
\usepackage{array}
\usepackage{latex8}
\usepackage{babel}
\usepackage{tabularx}
\usepackage{longtable}
\usepackage{color}
\usepackage{graphicx}
\usepackage{overpic}

\makeatletter

\newcommand{\mypara} [1] {\smallskip{} {\bf #1}}

\begin{document}

\title{Machine Learning and Cloud Computing: Survey of Distributed and SaaS Solutions \thanks{This manuscript was originally published as IEAT Technical Report at https://www.ieat.ro/technical-reports in 2012.}}

\author{Daniel Pop}

\affiliation{Institute e-Austria Timi\c{s}oara \\
Bd. Vasile P\^{a}rvan No. 4, 300223  Timi\c{s}oara, Rom\^{a}nia}

\email{danielpop@info.uvt.ro}

\maketitle

\begin{abstract}
Applying popular machine learning algorithms to large amounts of data raised new challenges for the ML practitioners. Traditional ML libraries does not support well processing of huge datasets, so that new approaches were needed. Parallelization using modern parallel computing frameworks, such as MapReduce, CUDA, or Dryad gained in popularity and acceptance, resulting in new ML libraries developed on top of these frameworks. We will briefly introduce the most prominent industrial and academic outcomes, such as Apache Mahout\texttrademark, GraphLab or Jubatus. 
		
We will investigate how cloud computing paradigm impacted the field of ML. First direction is of popular statistics tools and libraries (R system, Python) deployed in the cloud. A second line of products is augmenting existing tools with plugins that allow users to create a Hadoop cluster in the cloud and run jobs on it. Next on the list are libraries of distributed implementations for ML algorithms, and on-premise deployments of complex systems for data analytics and data mining. Last approach on the radar of this survey is ML as Software-as-a-Service, several BigData start-ups (and large companies as well) already opening their solutions to the market.
\end{abstract}

\section{Introduction}

Given the enormous growth of collected and available data in companies, industry and science, techniques for analyzing such data are becoming ever more important.  Today, data to be analyzed are no longer restricted to sensor data and classical databases, but more and more include textual documents and webpages (text mining, Web mining), spatial data, multimedia data, relational data (molecules, social networks). Analytics tools allow end-users to harvest the meaningful patterns buried in large volumes of structured and unstructured data. Analyzing big datasets gives users the power to identify new revenue sources, develop loyal and profitable customer relationships, and run your overall organization more efficiently and cost effectively.

Research in knowledge discovery and machine learning combines classical questions of computer science (efficient algorithms, software systems, databases) with elements from artificial intelligence and statistics up to user oriented issues (visualization, interactive mining). 

\begin{figure}
\includegraphics[width=0.48\textwidth]{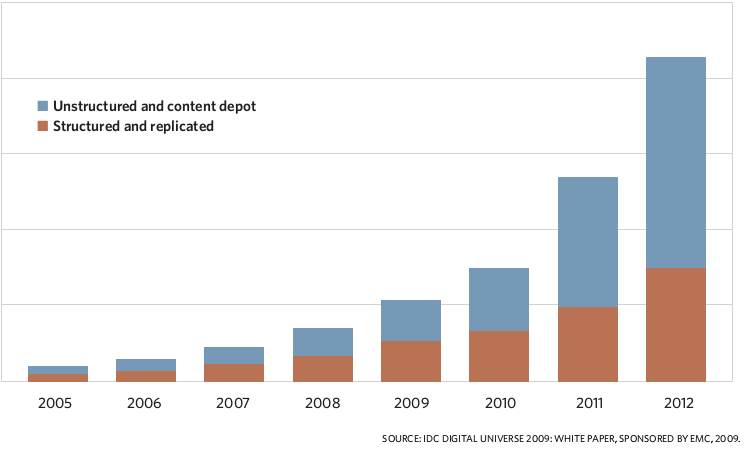}
\caption{Trends in data growth}
\label{figure:data-growth}
\end{figure}

Although for more than two decades, parallel database products, such as Teradata, Oracle or Netezza have provided means to realize a parallel implementation of ML-DM algorithms, expressing ML-DM algorithms in SQL code is a complex task and difficult to maintain. Furthermore, large-scale installations of these products are expensive and are not an affordable option in most cases. Another driver for paradigm shift from relational model to other alternatives is the new nature of data. Until about five years ago, most data was transactional in nature, consisting of numeric or string data that fit easily into rows and columns of relational databases. Since then, while structured data is following a near-linear growth, unstructured (e.g. audio and video) and semi-structured data (e.g, Web traffic data, social media content, sensor generated data etc.) exhibit an exponential growth (see figure~\ref{figure:data-growth}). Most of the new data is either semi-structured in format, i.e. it consists of headers followed by text strings, or pure unstructured data (photo, video, audio). While the latter has limited textual content and is more difficult to parse and analyze, semi-structured data triggered a plethora of non-relational data stores (NoSQL data stores) solutions tailored to handle huge amount of data. Consequently, the past 5 years have seen researchers moving to parallelization of ML-DM using these new platforms, such as NoSQL datastores, distributed processing environments (MapReduce), or cloud computing.

At this point, it is worth reflecting to a nice metaphor by Ben Werther~\cite{Werther-12}, co-founder of Platfora, for big data processing today: 

\medskip{}
\begin{overpic}[scale=.6]{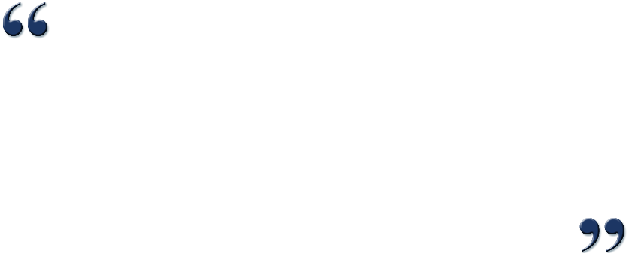}
\put(10,35){\textit{In ‘industrial revolution’ terms, we are in}}
\put(10,28){\textit{the pre-industrial era of artisanship that}}
\put(10,21){\textit{preceeded mass production.}}
\put(10,14){\textit{It is the equivalent of needing to engage an}}
\put(10,7){\textit{expert blacksmith to forge the forks and}}
\put(10,0){\textit{spoons for our dinner table.}}
\end{overpic}
\smallskip{}

Machine Learning is inherently a time consuming task, thus plenty of efforts were conducted to speed-up the execution time. Cloud computing paradigm and cloud providers turned out to be valuable alternatives to speed-up machine learning platforms. Thus, popular statistics tools environments – like R, Octave, Python – went in the cloud as well. There are two main directions to integrate them with cloud providers: create a cluster in the cloud and bootstrapping it with statistic tools, or augment statistic environments with plugins that allow users to create Hadoop clusters in the cloud and run jobs on them.

Environments like R, Octave, Mapple and similar offer low-level infrastructure for data analysis, that can be applied for large datasets once leveraged by cloud providers. Machine Learning is something that comes on top of this and facilitates the retrieval of useful knowledge out of huge data for customers with no/less statistical background by automatically inferring ‘knowledge models’ out of data. To support this need, an explosion of start-ups, some of them in stealth mode yet, who are {\em offering machine learning services} to their customers, or big data analysis services can be noticed in past 5 years. These initiatives can be either PaaS/SaaS platforms or products that can be deployed on private environments.

\bigskip{}

Reviewing the literature and the market, we can conclude that ML-DM comes in many flavors. We classify these approaches in 5 distinct classes:
\begin{itemize}
\item Machine Learning environments from the cloud -- create a computer cluster in the cloud and bootstrapping it with statistics tools. $\Rightarrow$ Section~\ref{sec: ML-clouds}.

\item Plugins for Machine Learning tools -- augment statistics tools with plugins that allow users to create a Hadoop cluster in the cloud and run ML jobs on it.  $\Rightarrow$ Section~\ref{sec: ML-cloud-plugins}.

\item Distributed Machine Learning libraries -- collections of parallelized implementations of ML algorithms for distributed environments (Hadoop, Dryad etc). $\Rightarrow$ Section~\ref{sec: Distributed-ML-Frameworks}.

\item Complex Machine Learning systems -- products that need to be installed on private data centers (or in the cloud) and offers high performance data mining and analysis. $\Rightarrow$ Section~\ref{sec: ML-tools}.

\item Software as a Service providers for Machine Learning -- PaaS/SaaS solutions that allow clients to access ML algorithms via Web services.  $\Rightarrow$ Section~\ref{sec: SaaS-Providers-for-ML}.
\end{itemize}

The remaining of the paper is structured as follows: next section presents similar, recent studies, followed by 5 sections, each of them devoted to a particular class identified above. The paper ends with conclusion and future plans. 

\section{\label{sec: Previous-Work}Related studies}

Since 1995, many implementations were proposed for ML-DM algorithms parallelization for shared or distributed systems. For a comprehensive study the reader is referred to a recent survey~\cite{Upadhyaya-13}. Our work is focused in frameworks, toolkits, libraries that allow large-scale, distributed implementations of state-of-the-art ML-DM algorithms. To this respect, we mention a recent book dealing with machine learning at large~\cite{Bekkerman-12}, which contains both presentations of general frameworks for highly scalable ML implementations, like DryadLINQ or IBM PMLT, and specific implementations of ML techniques on these platforms, like ensemble decision trees, SVM, k-Means etc. It contains contributions from both industry leaders (Google, HP, IBM, Microsoft) and academia (Berkeley, NYU, University of California etc).

Recent articles, such as those of S. Charrington~\cite{Charrington-12}, W. Eckerson~\cite{Eckerson-12} and D. Harris~\cite{Harris-12}, review different large-scale ML solutions providers that are trying to offer better tools and technologies, most of them based on Hadoop infrastructure, to move forward the novel industry of big data. They are aiming at improving user experience, at product recommendations, or website optimization applicable for finance, telecommunications, retail, advertising or media. 

\section{\label{sec: ML-clouds}Machine Learning environments from the cloud}

Providers of this category offer computer clusters using public cloud providers, such as Amazon EC2, Rackspace etc, pre-installed with statistics software, preferred packages being R system, Octave or Mapple. These solutions offer scalable high-performance resources in the cloud to their customers, who are freed from the burden of installating and managing own clusters.

\mypara{Cloudnumbers.com} \footnote{http://cloudnumbers.com} are using Amazon EC2~\footnote{http://aws.amazon.com/ec2/} provider to setup computer clusters preinstalled with software for scientific computing, such as R system, Octave or Mapple. Customers benefit from a web-interface where they can create own workspaces, configure and monitor the cluster, upload datasets or connect to public databases. On top of default features from cloud provider, Cloudnumbers offers high security standards by providing secure encryption for data transmission and storage. Overall, a HPC platform in the cloud, easy to create and effortless to maintain. 

\mypara{CloudStat}~\footnote{http://cs.croakun.com} is a cloud integrated development environment built based on R system, and exposes its functionalities via 2 types of user interfaces: console -- for experienced users in R language, and applications -- designed as a point and click forms based interface for R for users with no R programming skills. There is also a CloudStat AppStore where users can choose applications from a growing repository.

\mypara{Opani}~\footnote{http://opani.com} is offering similar services to Cloudnumbers.com, but additionally helps customers to size their cluster according to their needs: size of data and the time-frame for processing this data. They are using Rackspace's~\footnote{http://rackspace.com} infrastructure and support environments such as R system, Node and Python, bundled with map/reduce, visualization, security and version control packages. Results of data analysis processes, named dashboard in Opani, can easily be visualized and shared from desktop or mobile devices.

\smallskip{}

Approaches in this class are powerful and flexible solutions, offering users the possibility to develop complex ML-DM applications ran on the cloud. Users are freed from the burden of provisioning own distributed environments for scientific computing, while being able to use their favorite environments. On the other side, users of these tools need to have extensive experience in programming and strong knowledge of statistics. Perhaps, due to this limited audience, the stable providers in this category are fewer than in other categories, some of them (such as CRdata.org) shutting down the operation only shortly after taking off.

\section{\label{sec: ML-cloud-plugins}Plugins for Machine Learning toosl}

In this class, statistics applications (e.g. R system, Python) are extended with plugins that allow users to create a Hadoop cluster in the cloud and run time consuming jobs over large datasets on it. Most of the interest went towards R, for which several extensions are available, comparing to Python for which less effort was invested until recently in supporting distributed processing. In this section we will mention several solutions for R and Python.

\mypara{RHIPE}~\footnote{http://www.stat.purdue.edu/~sguha/rhipe/doc/html/index.html} is a R package that implements a map/reduce framework for R offering access to a Hadoop installation from within R environment. Using specific R functions, users are able to launch map/reduce jobs executed on the Hadoop cluster and results are then retrieved from HDFS.

\mypara{Snow}~\footnote{http://cran.r-project.org/web/packages/\\available\textunderscore{}packages\textunderscore{}by\textunderscore{}name.html}~\cite{Tierney-09} and its variants (snowfall, snowFT) implement a framework that is able to express an important class of parallel computations and is easy to use within an interactive environment like R. It supports three types of clusters: socket-based, MPI, and PVM.

\mypara{Segue for R}~\footnote{http://code.google.com/p/segue/} project makes it easier to run map/reduce jobs from within R environment on elastic clusters at Amazon Elastic Map Reduce~\footnote{http://aws.amazon.com/elasticmapreduce/}. 

\mypara{Anaconda}~\footnote{https://store.continuum.io/cshop/anaconda} is a scalable data analytics and scientific computing in Python offered by Continuum Analytics~\footnote{http://continuum.io}. It is a collection of packages (NumbaPro -- fast, multi-core and GPU-enabled computation, IOPro -- fast data access, and wiseRF Pine -- multi-core implementation of the Random Forest) that enables large-scale data management, analysis, and visualization and more. It can be installed as a full Python distribution or can be plugged into an existing installation.

\smallskip{}

Due to its popularity among ML-DM practitioners, R system being the preferred tool for such tasks in past 2 years \cite{rexer-11, kdnuggets-12}, efforts have been made recently to parallelize lengthy processes on scalable distributed frameworks (Hadoop). This approach is largely preferred over ML in the cloud due to the possibility to re-use existing infrastructure of research, or industrial (private) data centers. To the best of our knowledge, there are no similar approaches for related mathematical tools, such as Mathematica, Maple or Matlab/Octave, except HadoopLink~\footnote{https://github.com/shadanan/HadoopLink} for Mathematica. The audience of this class of solutions is also highly qualified in programming languages, mathematics, statistics and machine learning algorithms.

\section{\label{sec: Distributed-ML-Frameworks}Distributed Machine Learning libraries}

This category offers complex libraries operating on various distributed setups (Hadoop, Dryad, MPI). They allow users to use out-of-the-box algorithms, or implement their own, that are run in parallel mode over a cluster of computers. These solutions does not integrate, nor use, statistics/mathematics software, rather they offer self-contained packages of optimised, state-of-the-art ML-DM methods and algorithms.

\mypara{Apache Mahout\texttrademark}~\footnote{http://mahout.apache.org}~\cite{Owen-11} is an Apache project to produce free implementations of distributed or otherwise scalable machine learning algorithms on the Hadoop platform~\cite{Hadoop-web}.  It started as a collection of independent, "Hadoop-free" components, e.g. "Taste" collaborative-filtering. Its goal is to build \textit{scalable} machine learning libraries, where scalable has a broader meaning:
\begin{itemize}
\item Scalable to reasonably large datasets. Mahout's core algorithms for clustering, classification and batch based collaborative filtering are implemented on top of Apache Hadoop~\cite{Hadoop-web} using the map/reduce paradigm. However, it does not restrict contributions to Hadoop based implementations: contributions that run on a single node or on a non-Hadoop cluster are welcome as well. The core libraries are highly optimized to allow for good performance also for non-distributed algorithms.
\item Scalable to support various business cases. Mahout is distributed under a commercially friendly Apache Software license. 
\item Scalable community. The goal of Mahout is to build a vibrant, responsive, diverse community to facilitate discussions not only on the project itself but also on potential use cases.
\end{itemize}

Currently Mahout supports mainly four use cases: 
\begin{itemize}
\item Recommendation mining takes users' behavior and from that tries to find items users might like
\item Clustering takes e.g. text documents and groups them into groups of topically related documents
\item Classification learns from existing categorized documents what documents of a specific category look like and is able to assign unlabelled documents to the (hopefully) correct category. 
\item Frequent itemset mining takes a set of item groups (terms in a query session, shopping cart content) and identifies, which individual items usually appear together.
\end{itemize}

Integration with initiatives such as graph processing platforms Apache Giraph~\footnote{http://incubator.apache.org/giraph/} are actively under discussion. An active community is behind this project.

\mypara{GraphLab}~\footnote{http://graphlab.org}~\cite{Low-12} is a framework for ML-DM in the Cloud.
While high-level data parallel frameworks, like MapReduce, simplify the design and implementation of large-scale data processing systems, they do not naturally or efficiently support many important data mining and machine learning algorithms and can lead to inefficient learning systems. To help fill this critical void, GraphLab is an abstraction which naturally expresses asynchronous, dynamic, graph-parallel computation while ensuring data consistency and achieving a high degree of parallel performance, in both shared-memory and distributed settings. It is written in C++ and is able to directly access data from Hadoop Distributed File System (HDFS)~\cite{Hadoop-web}. The authors report out-performing similar approaches by orders of magnitude.

\mypara{DryadLINQ}~\footnote{http://research.microsoft.com/en-us/projects/DryadLINQ/}~\cite{Yu-08, Budiu-12} is LINQ (Language INtegrated Query~\footnote{http://msdn.microsoft.com/netframework/future/linq/} subsystem developed at Microsoft Research on top of Dryad~\cite{Isard-07}, a general purpose architecture for execution of data parallel applications. It supports DAG-based abstractions, inherited from Dryad, for implementing data processing algorithms.  A DryadLINQ program is a sequential program composed of LINQ expressions performing arbitrary side-effect-free transformations on datasets, and can be written and debugged using standard .NET development tools. The DryadLINQ system automatically and transparently translates the data-parallel portions of the program into a distributed execution plan which is passed to the Dryad execution platform that ensures efficient and reliable execution of this plan. Authors demonstrate near-linear scaling of execution time on the number of computers used for a job. While the DAG-based abstraction permits rich computational dependencies, it does not naturally express iterative, data parallel, task parallel and dynamic data driven algorithms that are prevalent in ML-DM.

\mypara{Jubatus}~\footnote{http://jubat.us/}~\cite{Hido-12}, started April 2011, is an online/real-time machine learning platform, implemented on a distributed architecture. Comparing to Mahout\texttrademark is a next-step platform that offers stream processing and online learning. In online ML, the model is continuously updated with each data sample that is coming by fast and not memory-intensive algorithms. It requires no data storage, nor sharing; only model mixing.  It supports classification problems (Passive Aggressive (PA), Confidence Weighted Learning, AROW), PA-based regression, nearest neighbor (LSH, MinHash, Euclid LSH), recommendation, anomaly detection (LOF based on NN) and graph analysis (shortest path, PageRank).  In order to efficiently support online learning, Jubatus operates updates on local models and then each server transmits its model difference that are merged and distributed back to all servers. The mixed model improves gradually thanks to all servers' work. 

\mypara{IBM Parallel Machine Learning Toolbox}~\footnote{https://www.research.ibm.com/haifa/projects/verification/\\ml\textunderscore{}toolbox/index.html}~\cite{Pednault-12} (PMLT), a joint effort of the Machine Learning group at the IBM Haifa Lab and the Data Analytics department at the IBM Watson Lab, provides tools for execution of data mining and machine learning algorithms on multiple processor environments or on multiple threaded machines.  The toolbox comprises two main components: an API for running the users' own machine learning algorithms, and several pre-programmed algorithms which serve both as examples and for comparison. The pre-programmed algorithms include a parallel version of the Support Vector Machine (SVM) classifier, linear regression, transform regression, nearest neighbors, k-means, fuzzy k-means, kernel k-means, PCA, and kernel PCA. One of the main advantages of the PML toolbox is the ability to run it on a variety of operating systems and platforms, from multi-core laptops to supercomputers such as BlueGene. This is because the toolbox incorporates a parallelization infrastructure that completely separates parallel communications, control, and data access from learning algorithm implementation. This approach enables learning algorithm designers to focus on algorithmic issues without having to concern themselves with low-level parallelization issues. It also enables learning algorithms to be deployed on multiple hardware architectures, running either serially or in parallel, without having to change any algorithmic code. The toolbox uses the popular MPI library as the basis for its operation, and is written in C++. Despite of our effort to get latest news on this project, we found no recent activity on this project since 2007, except for a chapter in~\cite{Bekkerman-12} (2012). On the other side, the toolkit is suited for parallel environments, not for distributed ones.

\mypara{NIMBLE}~\cite{Ghoting-11} is a sequel project to Parallel Machine Learning Toolbox, also developed at IBM Research Labs. It exposes a multi-layered framework where developers may express their ML-DM algorithms as tasks. Tasks are then passed to the next layer, an architecture independent layer, composed of one queue of DAGs of tasks, plus worker threads pool that unfold this queue. Next layer is an architecture dependent layer that translates the generic entities from upper layer into various runtimes. Currently, NIMBLE supports execution on Hadoop platform~\cite{Hadoop-web} only. Other platforms, such as Dryad~\cite{Isard-07}, are also good candidates, but not yet supported. Advantages of this framework include:
\begin{itemize}
\item higher level of abstraction, hiding low-level control and choreography details of most of the distributed and parallel programming paradigms (MR, MPI etc), allowing programmers to compose parallel ML-DM algorithms using reusable (serial and parallel) building blocks
\item portability: providing specific implementation for architecture dependent layer, same code can be executed on various distributed runtimes
\item efficiency and scalability: due to optimisation introduced by DAGs of tasks and co-scheduling, results presented in~\cite{Ghoting-11} for Hadoop runtime show speedup improvement with increasing dataset size and dimensionality.
\end{itemize}

\mypara{SystemML}~\cite{Ghoting-11-2}, developed at IBM Research labs as NIMBLE and PMLT, proposes an R-like language (Declarative Machine Learning language) that includes linear algebra primitives and shows how it can be optimized and compiled down to MapReduce. They report an extensive performance evaluation on three (Group Nonnegative Matrix Factorization, Liner regression, Page Rank) ML algorithms on varying data and Hadoop cluster sizes.

\smallskip{}

Table \ref{table: distributed-frameworks-ml-dm} presents a synthesis on investigated platforms. One can notice that Java is the preferred environment, due to large adoption and usage of Hadoop as distributed processing model. The good news is the fact that most active and lively solutions are the open-source ones. Target audience of this class of products are programmers, system developers and ML experts who need fast, scalable distributed solutions for ML-DM problems.

\begin{table*}[h]
\begin{tabular}{ccccc}
\hline 
Name  & Platform & Licensing  & Language & Activity \\ 
\hline 
\hline 
Mahout & Hadoop & Apache 2 & Java & High \\ \hline
GraphLab & MPI / Hadoop &  Apache 2 & C++ & High \\ \hline
DryadLINQ & Dryad & Commercial & .NET & Low \\ \hline
Jubatus & ZooKeeper & LGPL 2 & C++ & Medium \\ \hline
NIMBLE & Hadoop & ? & Java & Low \\ \hline
SystemML & Hadoop & ? & DML & Low \\ \hline
\end{tabular}
\label{table: distributed-frameworks-ml-dm}\caption{Distributed Frameworks for ML-DM}
\end{table*}

\section{\label{sec: ML-tools}Complex Machine Learning systems}

This section present several solutions for business intelligence and data analytics that share a set of common features: (i) all are deployable on on-premise or in-the-cloud clusters, (ii) provide rich set of graphical tools to analyse, explore and visualize large amounts of data, (iii) expose a rather limited set of ML-DM functions, usually limited to prediction models and (iv) utilize Apache Hadoop~\cite{Hadoop-web} as processing engine and/or storage environment. There are differences on how data is integrated and processed, supported data sources or related to complexity of the system. Here are the most known ones:

\mypara{Kitenga Analytics}~\footnote{http://www.quest.com/news-release/quest-software-expands-its-big-data-solution-with-new-hadoop-ce-102012-818658.aspx}, recently purchased by Dell, is a native Hadoop application that offers visual ETL, Apache Solr\texttrademark~\footnote{http://lucene.apache.org/solr/}-based search, natural language processing, Apache Mahout-based data mining, and advanced visualization capabilities. It is a big data environment for sophisticated analysts who want a robust toolbox of analytical tools, all from an easy-to-use interface that that does not require understanding of complex programming or the Apache Hadoop stack itself.

\mypara{Pentaho Business Analytics}~\footnote{http://www.pentaho.org} offers a complete solution for big data analytics, supporting all phases of an analytics process, from pre-processing to advanced data exploration and visualization. It offers (i) a complete visual design tool to accelerate data preparation and modeling, (ii) data integration from NoSQL and relational databases, (iii) distributed execution on Hadoop platform~\cite{Hadoop-web}, (iv) instant and interactive analysis (no code, no ETL (Extract, Transform, Load)) and (v) business analytics platform: data discovery, exploration, visualization and predictive analytics. Main characteristics of Pentaho solution include:
\begin{itemize}
\item MapReduce-based data processing
\item Can be configured for different Hadoop distributions (such as Cloudera, Hadapt etc.)
\item Data can be loaded and processed into Hadoop HDFS, HBase~\footnote{http://hbase.apache.org}, or Hive~\footnote{hive.apache.org}
\item Supports Pig scripts
\item Native support for most NoSQL databases, such as Apache Cassandra, DataStax, Apache HBase, MongoDB, 10gen etc.
\item Enables performance-optimized data analysis, reporting and data integration for analytic databases (such as Teradata, monetdb, Netezza etc.), through deep integration with native SQL dialects and parallel bulk data loader
\item Integration wit HPCC (High Performance Computing Cluster) from LexisNexis Risk Solutions~\footnote{http://hpccsystems.com}
\item Import/export from/to PMML (Predictive Modeling Markup Language)
\item Pentaho Instaview, a visual application to reduce the time needed to deploy data analytics solutions and to help novice users to get insights of their data, in three simple steps: select data source, automatically prepare data for analytics, and visualize and explore built models.
\item Pentaho Mobile - application for iPad that provides interactive business analytics for business users
\end{itemize}

Their ecosystem is composed of several powerful systems, each of them a complex project of its own:

{\em Pentaho BI Platform/Server}  the BI platform is a framework providing core services, such as authentication, logging, auditing and rules engines; it also has a solution engine that integrates all other systems (reporting, analysis, integration and data mining); BI Server is the most well known implementation of the platform, which functions as a web based report management system, application integration server and lightweight workflow engine.

{\em Pentaho Reporting} based on JFreeReport, is a suite of open-source tools -- Pentaho Report Designer, Pentaho Reporting Engine, Pentaho Reporting SDK and the common reporting libraries shared with the entire Pentaho BI Platform -- that allows users to create relational and analytical reports from a variety of sources outputting results in various formats (HTML, PDF, Excel etc.)

{\em Pentaho Data Integration (Kettle)} delivers powerful ETL capabilities using metadata-driven approach with an intuitive, graphical, drag and drop design environment; 

{\em Pentaho Analysis Service (Mondrian)} is an Online Analytical Processing (OLAP) server that supports data analysis in real-time

{\em Pentaho Data Mining (Weka)} a collection of machine learning algorithms for classification, regression, clustering and association rules;

\mypara{Platfora}~\footnote{http://platfora.com} delivers in-memory business intelligence with no separate data warehouse or ETL required. Its visual interface built on HTML5 allows business users to analyse data. Results may be easily shared between users. It relies on Hadoop cluster, that can be installed either on own premise, or on cloud providers (Amazon EMR and S3). It is primarly focused on BI features, such as elaborated visualization types (charts, plots, maps), or slice-and-dice operations, but also offers a predictive analysis framework.

\mypara{Skytree Server}~\footnote{http://skytree.net} is a general purpose machine learning and data analytics system that supports data coming from relational databases, Hadoop systems, or flat files and offers connectors to common statistical packages and ML libraries. ML methods supported are: Support Vector Machine (SVM), Nearest Neighbor, K-Means, Principal Component Analysis (PCA), Linear Regression, 2-point correlation and Kernel Density Estimation (KDE). Skytree Server connects with analytics front-ends, such as Web services or statistical and ML libraries (R, Weka), for data visualization. Its deployment options include cloud providers, or dedicated cluster based on Linux machines. It also supports customers in estimating the size of the cluster they need by a simple formula (Analytics Requirements Index).

\mypara{Wibidata}~\footnote{http://wibidata.com} is a complex solution based on open source software stack from Apache, combining Hadoop, HBase and Avro with proprietary components. WibiData's machine learning libraries give the tools to start building complex data processing pipelines immediately. WibiData also provides graphical tools to export your data from its distributed data repository into any relational database \cite{Wibidata-how-it-works}. In order to simplify data processing using Hadoop, WibiData introduces the concepts of {\em producers} -- computation functions that update a row in a table, and {\em gatherers} -- close the gap between WibiData table and key-value pairs processed by Hadoop MapReduce engine.

\smallskip{}

We are aware that we could not cover all the solution provider in the field of business intelligence and big data analytics. We tried to cover those who are also offering ML components in their applications, many others focusing only on big data analytics, such as Alteryx, SiSense, SAS or SAP, being omitted from this survey. Solutions in this category target mostly business users, who need to quickly and easily extract insights from their data, being good candidates for users with less computer or statistics background.

\section{\label{sec: SaaS-Providers-for-ML}Software as a Service providers for Machine Learning}

This section focuses on platform-as-a-service, or software-as-a-service providers for machine learning problems. They are offering the services mainly via RESTful interfaces, and in some (rare) cases the solution may also be installed on-premise (Myrrix), contrasting to solutions from previous section that are mainly deployable systems on private data centers. As class of ML problems, predictive modeling is the favorite (BigML, Google Prediction API, Eigendog) among these systems. We did not include in this study providers of SQL over Hadoop solutions (e.g. Cloudera Impala, Hadapt, Hive) because their main target is not ML-DM, rather fast, elastic and scalable SQL processing of relational data using the distributed architecture of Hadoop.

\mypara{BigML}~\footnote{http://bigml.com} is a SaaS approach to machine learning. Users can setup datasources, create, visualize and share prediction models (only decision trees are supported), and use models to generate predictions. All from a Web interface or programmatically using REST API. 

\mypara{BitYota}~\footnote{http://bityota.com} is a young start-up (2012) SaaS provider for BigData warehousing solution. On top of data integration from different sources (relational, NoSQL, HDFS) it also allows customers to run statistics and summarization queries in SQL92, standard R statistics and custom functions written in JavaScript, Perl, or Python on a parallel analytics engine. Results are visualized by integrating with popular BI tools and dashboards.

\mypara{Precog}~\footnote{http://precog.com} has a more elaborate SaaS solution composed of Precog database, Quirrel language, ReportGrid and LabCoat tools. At the core of Precog, we have an original (no Hadoop, no other NoSQL based), schemaless, columnar database designed for storing and analyzing semi-structured, measured data, such as events (users clicking, engaging, and buying), sensor data, activity stream data, facts, and other kinds of data that do not need to be mutably updated. Precog’s functionality is exposed by REST APIs, but client libraries are available in JavaScript, Python, PHP, Ruby, Java, or C\#. LabCoat is a GUI tool for creation and management of Quirrel queries. Quirrel is a a highly expressive data analysis language that makes it easy to do in-database analytics, statistics, and machine learning across any kind of measured data. Results are available in JSON or CSV formats. ReportGrid is an HTML5 visualization engine that interactively, or programmatically, build reports and charts. 

\mypara{Google Prediction API}~\footnote{https://developers.google.com/prediction/} is Google's cloud-based machine learning tools that can help analyze your data. It is closely connected to Google Cloud Storage\footnote{https://developers.google.com/storage/} where training data is stored and offers its services using a RESTful interface, client libraries allowing programmers to connect from Java, JavaScript, .NET, Ruby, Python etc. In the first step, the model need to be trained from data, supported models being classification and regression for now. After the model is built, one can query this model to obtain predictions on new instances. Adding new data to a trained model is called Streaming Training and it is also nicely supported. Recently, PMML preprocessing feature has been added, i.e. Prediction API .supports preprocessing your data against a PMML transform specified using PMML 4.0 syntax; does not support importing of a complete PMML model that includes data. Created models can be shared as hosted models in the marketplace. 

\mypara{EigenDog}~\footnote{https://eigendog.com/\#home} is a service for scalable predictive modeling, hosted on Amazon EC2 (for computation) and S3 (for data and models storage) platforms. It builds decision tree model out of data in Weka's ARFF format. Models can be downloaded in binary format and integrated in user applications thanks to API, or open-source library provided by vendor.

\mypara{Metamarkets}~\footnote{http://metamarkets.com/} claim as being Data Science-as-a-Service providers, helping users to get insights out of their large datasets. They offer end-users the possibility to perform fast, ad-hoc investigations on data, to discover new and unique anomalies, to spot trends in data streams, based on statistical models, in an intuitive, interactive and collborative way. They are focused on business people, less knowledgeable on statistics and machine learning.

\mypara{Myrrix}~\footnote{http://myrrix.com} is a complete, real-time, scalable recommender system built using Apache Mahout\texttrademark (see Section~\ref{sec: Distributed-ML-Frameworks}). It can be accessed as PaaS using a RESTful interface. It is able to incrementally update the model once new data is available. It is organized in 2 layers -- Serving (open source and free) and Computation (Hadoop based) -- that can be deployed on-premise as well, either both of them or only one.

\mypara{Prior Knowledge Veritable API}~\footnote{http://priorknowledge.com} offers Python and Ruby interfaces; upload data on their servers, and build prediction model using Markov Chain Monte Carlo samplers. They were operating a cloud based infrastructure based on Amazon WS. SalesForce.com acquired Prior Knowledge at the end of 2012.

\mypara{Predictobot}~\footnote{http://predictobot.com} by Prediction Appliance also aims at doing machine learning modeling easier. The user will upload a spreadsheet of data, answer a few questions, and then download a spreadsheet with the predictive model.  It is going to bring predictive modeling to anyone with the skills to make a spreadsheet.  The business is still in stealth mode.

\subsection{Text mining as SaaS}
Due to explosion of social media technologies, such as blog platforms (WordPress.com, Blogger etc), mini-blogging (Twitter), or social networks (Facebook, Google+), an increased interest is paid to text mining and natural language processing (NLP) solutions delivered as services to their customers. This is why we devoted an entire subsection to group together software/platform-as-a-service solutions for text mining. Before reviewing available solutions, a short introduction to NLP and text mining is helpful.

While NLP uses linguistically inspired techniques (text is syntactically parsed using information from a formal grammar and a lexicon, and the resulting information is then interpreted semantically and used to extract information) to deeply analyse the document, text mining is more recent and uses techniques developed in the fields of information retrieval, statistics, and machine learning. Contrasting with NLP, text mining's aim is not to understand what is "said" in a text, rather to extract patterns across large number of documents. Features of text mining include extraction of concept/entity, text clustering, summarization, or sentiment analysis.

Size and number of documents that need to be processed, plus real-time processing constrain contribute to the development of novel, distributed toolkits able to answer demanding users' needs. Websites operators are willing to offer text mining features to their visitors with minimum investment and reduced maintenance costs. Thus, more and more providers are offering text mining services through RESTful web services, saving clients from costly infrastructures and deployments.

Without aiming at providing an exhaustive survey of text mining P(S)aaS providers, we will mention several of them hereafter:

\mypara{AlchemyAPI}~\footnote{http://www.alchemyapi.com} is a cloud-based text mining SaaS platform providing the most comprehensive set of NLP capabilities of any text mining platform, including: named entity extraction, sentiment analysis, concept tagging, author extraction, relations extraction, web page cleaning, language detection, keyword extraction, quotations extraction, intent mining, and topic categorization. AlchemyAPI uses deep linguistic parsing, statistical natural language processing, and machine learning to analyze your content, extracting semantic meta-data: information about people, places, companies, topics, languages, and more. It provides RESTful API endpoints, SDKs in all major programming languages and responses are encoded in various formats (XML, JSON, RDF). Organizations with specific data security needs or regulatory constraints are offered the possibility to install the solution on own environment.

\mypara{NathanApp\texttrademark}~\footnote{http://ai-one.com} is AI-one's general purpose machine learning PaaS, also available for deployment on-premise as NathanNode\texttrademark. Like Topic-Mapper, it is ideally suited to learn the meaning of any human language by learning the context of words, only faster and with greater deployment flexibility. NathanApp is a RESTful API using JavaScript and JSON.

\mypara{TextProcessing}~\footnote{http://text-processing.com} is also a NLP API that supports stemming and lemmatization, sentiment analysis, tagging and chunk extraction, phase extraction and named entity recognition. These services are offered open and free (for limited usage) via RESTful API endpoints, client libraries exist in Java, Python, Ruby, PHP and Objective-C, responses are JSON encoded and Python NLTK demos are offered to achieve a steep learning curve. For commercial purposes, clients are offered monthly subscriptions via Mashape.com.

\mypara{Yahoo! Content Analysis Web Service}~\footnote{http://developer.yahoo.com/search/content/V2/\\contentAnalysis.html} detects entities/concepts, categories, and relationships within unstructured content. It ranks those detected entities/concepts by their overall relevance, resolves those if possible into Wikipedia pages, and annotates tags with relevant meta-data. The service is available as an YQL table and response is in XML format. It is freely available for non-commercial usage.

\smallskip{}

This section presented PaaS solutions addressing, to some extent, machine learning problems. A special sub-section was devoted to text mining problem due to its spreading in the landscape of ML PaaS landscape. We notice big players, such as Yahoo! or Google, as well as many start-ups with million dollars fundings. They offer Web developers the possibility to easily integrate in their sites ML intelligence. Easy usage prevailed over functionality offered by these services, therefore there are only limited options of tweaking algorithms behind the services. Thus, these are good candidates for users with basic ML needs, but are not flexible enough for addressing more advanced problems.

\section{\label{sec:Conclusions}Conclusions and future work}

Our main findings are synthesized below:

(1) Existing programming paradigms for expressing large-scale parallelism such as MapReduce (MR) and the Message Passing Interface (MPI) are \textit{de facto} choices for implementing ML-DM algorithms. More and more interest has been devoted to MR due to its ability to handle large datasets and built-in resilience against failures.

(2) Machine Learning in distributed environments come in different approaches, offering viable and cost effective alternatives to traditional ML and statistical applications, which are not focused on distributed environments \cite{pop-11}.

(3) Existing solutions target either experienced, skilled computer scientists, mathematicians, statisticians or novice users who are happy with no (or few) possibilities to tune the algorithms. Ens-user support and guidance is largely missing from existing distributed ML-DM solutions.

After reviewing over 30 different offers on the market, we think that there is still room for a scalable, easy to use and deploy solution for ML-DM in the context of cloud computing paradigm, targeting end-users with less programming or statistical experience, but willing to run and tweak advanced scientific ML tasks, such as researchers and practitioners from fields like medicine, financial, telecommunications etc. To this respect, our future plans include prototyping such a distributed system relying on existing distributed ML-DM frameworks, but enhancing them with usability and user friendliness features.

\subsection*{Acknowledgments}

This work was supported by EC-FP7 project FP7-REGPOT-2011-1 284595 (HOST).


\pagebreak{}
{\bf Daniel Pop} received his PhD degree in computer science from West University of Timi\c{s}oara in 2006. He is currently a senior researcher at Department of Computer Science, Faculty of Mathematics and Computer Science, West University of Timi\c{s}oara. Research interests covers high performance computing and distributed computing technologies, machine learning and knowledge discovery and representation, and multi-agent systems. He also has a broad experience in IT industry (+15 years), where he applied agile software development processes, such as SCRUM and Kanban.

\end{document}